# Curvature-enhanced localised emission from dark states in wrinkled monolayer WSe$_2$ at room temperature

Sebastian Wood*, Filipe Richheimer, Tom Vincent, Vivian Tong, Alessandro Catanzaro, Yameng Cao, Olga Kazakova, and Fernando A. Castro

*National Physical Laboratory, Hampton Road, Teddington, Middlesex, TW11 0LW, UK*

*Corresponding author: sebastian.wood@npl.co.uk

**Abstract**

Localised emission from defect states in monolayer transition metal dichalcogenides is of great interest for optoelectronic and quantum device applications. Recent progress towards high temperature localised emission relies on the application of strain to induce highly confined excitonic states. Here we propose an alternative paradigm based on curvature, rather than in-plane stretching, achieved through free-standing wrinkles of monolayer tungsten diselenide (WSe$_2$). We probe these nanostructures using tip-enhanced optical spectroscopy to reveal the spatial localisation of out-of-plane polarised emission from the WSe$_2$ wrinkles. Based on the photoluminescence and Raman scattering signatures resolved with nanoscale spatial resolution, we propose the existence of a manifold of spin-forbidden excitonic states that are activated by the local curvature of the WSe$_2$. We are able to access these dark states through the out-of-plane polarised surface plasmon polariton resulting in enhanced strongly localised emission at room temperature, which is of potential interest for quantum technologies and photonic devices.

**Introduction**

Transition metal dichalcogenides (TMDs) have emerged as an interesting class of materials offering unique optoelectronic properties alongside the other advantages afforded by their 2D structure.[1-4] In particular, the Group 6 TMD monolayers are semiconductors exhibiting direct bandgaps in the visible range.[5] Specifically, monolayer WSe$_2$ (1L WSe$_2$) has emerged as a material of particular interest with observations of single photon emission from states with energies below that of the free bright exciton (A-exciton). Generally these sharp emission lines are only observed at cryogenic temperatures but it has been shown that strain-engineering can be used to confine excitons in deep potentials resulting in evidence of single-photon emission at temperatures as high as 150 K and 160 K.[6-8] Various strategies have been considered to tune the optoelectronic properties of TMDs through strain engineering in pursuit of efficient solid-state quantum light sources. These include the formation and use of wrinkles,[9-11] bubbles,[12-15] strained heterojunctions,[16] flexible substrates,[17,18] and structured substrates[6,7,19].

The precise identities of the emissive species responsible for localised emission in 1L WSe$_2$ remain unclear, and, whilst both local strain and defect states clearly have important roles in many of the published studies, the interplay between them is not fully resolved.[13,14,20] Uniaxial or biaxial tensile strain in 1L WSe$_2$ causes a local narrowing of the optical bandgap leading to confinement and the 'exciton funnelling' effect, where excitons drift towards the energetically favoured region of greatest strain resulting in enhanced and red-shifted local emission.[11,12,21-23] However, the energies of the single photon emitters are more consistent with defect states, and several groups have demonstrated that defects indeed play a role in the process.[7,24,25] A recent synthesis of these findings proposes that local strain lowers the energy of the 1L WSe$_2$ dark exciton resulting in hybridisation with an emissive point defect.[26]

The emerging consensus in the literature (cited above) is that the combination of applied strain with defects in 1L WSe$_2$ is an effective route to high temperature localised emission. However, most of the published research considers uniaxial and biaxial strains applied in the plane of the monolayer, which is not always the most appropriate approach. In fact, in the majority of published studies strain was applied in a way that also induces curvature, but the latter has been neglected from the analysis with the implicit failure to distinguish between in-plane strain and bending strain. The distinction between in-plane stretching strain and curvature (bending strain) is made graphically in Figure 1(a). In this study, we consider the case of a 'free-standing' wrinkle of 1L $WSe_2$ on a planar substrate, which induces curvature in the monolayer without (to first approximation) in-plane strain. The crest of the wrinkle has curvature like the bottom sketch in Figure 1(a), whereas the points of contact with the substrate exhibit an inverted curvature. Even if the finite thickness of the monolayer is considered, the tensile strain of the outer surface is balanced by compressive strain of the inner surface and hence the wrinkle is under bending strain but cannot be described in terms of pure in-plane strains. Rather we describe the wrinkle using curvature, understood as rarefaction of chalcogenide atoms on the outside of the curve and a compression of chalcogen atoms on the inside of the curve, whilst the transition metal atoms occupy a curved neutral plane. We would expect such a distortion of the atomic structure to result in tuning of the optoelectronic properties of the TMD in a distinct way from the application of in-plane strain. Clearly, for a material with finite thickness, it is not possible to create curvature without an associated bending strain field, but it is important to recognise that this is distinct from a pure in-plane stretching strain (which can be applied without creating curvature).

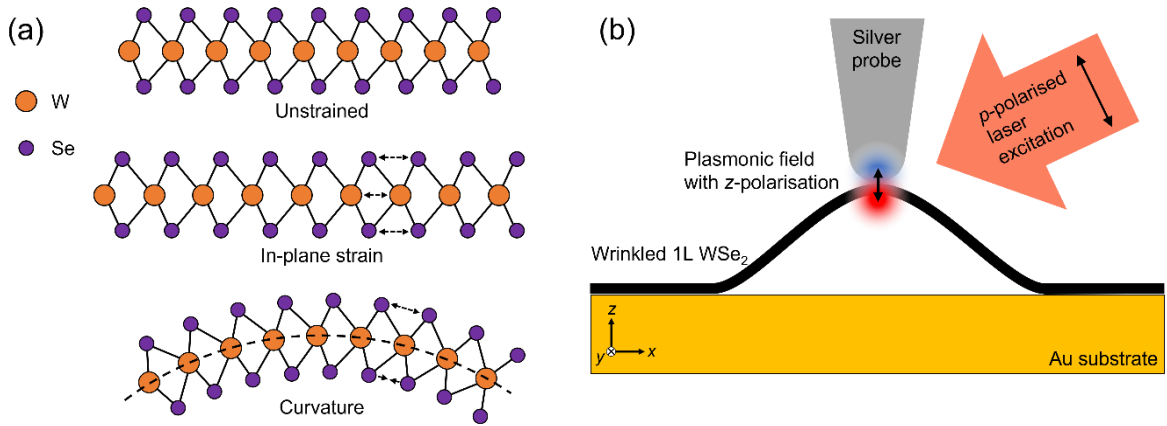

*Figure 1. (a) Diagram representing a cross-sectional structure of 1L WSe$_2$ comparing the unstrained structure with the effects of in-plane stretching strain and curvature (bending strain). Arrows show the changes in atomic spacings and the dashed line indicates the unstrained neutral plane in the curved structure. (b) Schematic diagram of the tip-enhanced optical spectroscopy measurement for wrinkled 1L WSe$_2$, showing the p-polarisation of the laser excitation and resulting z-polarisation of the plasmonic field.*

Since the free-standing wrinkles in 1L WSe$_2$ and the localised emissive sites are small with respect to optical wavelengths, we resolve the structure and emission using tip-enhanced optical spectroscopy (TEOS), where the near-field plasmonic coupling between a probe and the sample is used to achieve spectroscopy with nanoscale spatial resolution (Figure 1(b)).[27,28] TEOS is a particularly valuable technique for studying local optoelectronic and structural properties of 2D materials, where typical defects and features are highly localised to the length-scale 1 nm to 100 nm and so cannot be resolved with traditional far-field optical microspectroscopy.[11,14,29] The combination of atomic force microscopy (AFM) with a plasmonic probe is increasingly used across a range of disciplines where high resolution (typically ~ 10 nm) chemical or optical characterisation is required.[27,28,30-33] Two modes of TEOS are applied here: tip-enhanced photoluminescence (TEPL) and tip-enhanced Raman spectroscopy (TERS) to probe the localised emission and associated strain environment, respectively. An important feature of TEOS in the configuration used here is that the excitation laser is incident from the side with a *p*-polarisation (electric field approximately parallel to the AFM tip) to excite the *z*-polarised (out-of-plane) localised surface plasmon resonance supported by the metallised tip, resulting in strong coupling to out-of-plane transition dipoles in the WSe$_2$.[29,34]

In this study we demonstrate enhanced photoluminescence emission and Raman scattering signals from the crests of wrinkles in $WSe_2$. The emission includes contributions from localised low-energy emissive species that are not observed using conventional top-illumination far-field spectroscopy. These states are also not detected for the flat regions away from the wrinkles, indicating that curvature of the 1L $WSe_2$ (and the associated symmetry breaking) enhances the coupling of dark states to the out-of-plane polarised plasmon. Since spin-forbidden dark states are typically long-lived, this curvature-activated emission could be of particular value for quantum computing technologies.[34] Additionally, out-of-plane transition diploes are well-suited to on-chip photonic devices where light must be coupled into lateral waveguides.[13,25]

**Results**

The $WSe_2$ sample was prepared by mechanical exfoliation and deposited on an ultra-flat, template-stripped gold substrate to partially overlap a previously-transferred flake of few-layer graphene (FLG). Since the $WSe_2$/Au interface typically exhibits a van der Waals gap, which prevents efficient electronic interactions, the region with the interfacial flake of FLG was used as a control for this study, since it is expected that the $WSe_2$/FLG interface will exhibit strong quenching of the $WSe_2$ emission if the interface is clean.[35] The structure of the sample is indicated schematically in Figure 2 along with an optical micrograph and an atomic force microscopy (AFM) topography map. The FLG flake is a little over 10 μm across and shows some variation in thickness. The $WSe_2$ flake deposited on top of the FLG flake has a thick area near the middle of the FLG flake and a much thinner layer extending off the FLG at the top-left corner. This thin $WSe_2$ region has a wrinkled morphology resulting from the transfer process, which is the primary region of interest for this study.

Step-heights in topographic AFM do not always match the expected thicknesses of 2D materials due to differing adsorption of water and other species on the exposed surfaces so are not a robust method of evaluating layer thicknesses, nevertheless the FLG edge step height is ~1.1 nm and the $WSe_2$ flake step height is ~1.4 nm with respect to the Au substrate.[36] The optical contrast suggests that the graphene flake is best described as 'few layer' with a narrow stripe in the middle perhaps having only 1 or 2 layers. For the purposes of this study, the layer number of the graphene flake is not critical. The identification of part of the WSe$_2$ as a monolayer (1L) relies on the spectroscopic characterisation below. The sample is schematically represented in Figure 2(a) identifying the regions of 1L WSe$_2$, thicker WSe$_2$, and FLG. The outlines of the WSe$_2$ and FLG flakes are also shown on the optical micrograph in Figure 2(b).

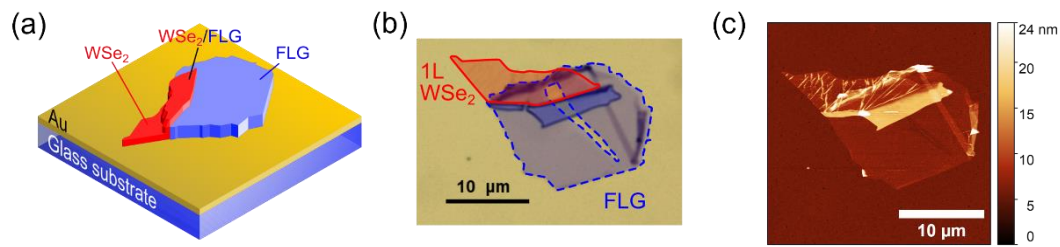

*Figure 2. (a) Schematic diagram showing WSe$_2$ flake (red) deposited over a few layer graphene (FLG) flake (blue) on a Au/glass substrate. (b) Optical microscope image of the sample with false colour identifying the FLG region (blue), and WSe$_2$ flake (red). (c) Topographic AFM image of the sample.*

The 1L WSe$_2$ flake exhibits a network of wrinkles, which are clearly visible in the topographic AFM (Figure 2(c)). A part of the WSe$_2$ monolayer with several wrinkles (1L WSe$_2$/Au) is shown in more detail in Figure 3(a). The wrinkles in this region are linear features with lengths up to 2 µm, which in some cases extend all the way across the flake. There is a dominant orientation of the wrinkles (bottom-left to top-right in the image) although not all the wrinkles are parallel. The form of these wrinkles is consistent with their formation during the transfer process. Deformation of the PDMS stamp results in an overall compressive strain in the deposited material sustained by adhesion between the 1L WSe$_2$ and the Au substrate surface. The compressive strain exceeds a threshold value resulting in buckle delamination to relieve in-plane strain.[23] As a result, the delaminated material maintains an energetic equilibrium of bending rigidity against interfacial adhesion with minimal residual in-plane strain.[37,38] Between the linear wrinkles there is a distribution of smaller point-features visible in the topographic map, which are similar to the 'nanobubbles' described elsewhere, arising from contaminants trapped under the WSe$_2$ (discussed in the Supporting Information).[14] The square area highlighted in Figure 3(a) was selected for detailed study of the structural and optical properties. The dimensions of the different wrinkles were examined by extracting cross-sectional profiles (Figure 3(c)) perpendicular to the wrinkle orientation for the positions of interest labelled B-F in Figure 3(b) (position A is a background point used for later comparisons). The cross-sectional profiles indicate that heights of these wrinkles are in the range 5 nm to 11 nm with respect to the flake. The widths of the wrinkles are harder to determine because they are convolved with the shape of the probe. Nevertheless, the apparent width of the wrinkles is 30 nm to 90 nm, which we may take as upper-bounds for the lateral dimension. These dimensions are comparable with those of WSe$_2$ wrinkles reported elsewhere.[9-11]

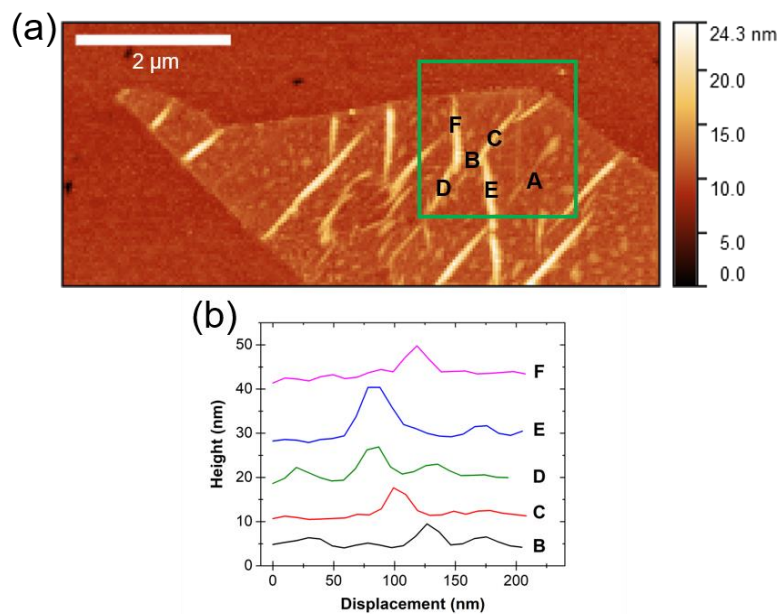

Figure 3. (a) Topographic AFM image of the monolayer WSe$_2$/Au region of the sample showing wrinkles and nanobubbles. Green square indicates region of interest for detailed study with specific positions labelled A-F. (c) AFM topography line profiles extracted perpendicular to wrinkles B-F, profiles are vertically offset for clarity.

Monolayer WSe$_2$ has a direct bandgap (in contrast to bilayer and multilayer WSe$_2$) with strong photoluminescence from the main bright exciton at around 1.66 eV.[14,34,39] Confocal photoluminescence microscopy with 1.96 eV excitation shows very strong emission from the area identified as 1L WSe$_2$/Au in Figure 4(a). The spatially averaged photoluminescence spectrum in Figure 4(b) shows a single peak centred at 1.656 eV, which is identified as free bright exciton emission. This strong emission also indicates that the 1L WSe$_2$ excitons are not strongly quenched by contact with the Au substrate, indicating a poor contact with the substrate due to trapped adsorbates or a van der Waals gap at the interface.[35] In contrast, where it overlaps with the FLG flake, the photoluminescence from the thin WSe$_2$ flake is strongly quenched (note the logarithmic scale in the photoluminescence map) and also exhibits a red-shift with the emission peak at 1.640 eV associated with efficient charge transfer at the WSe$_2$/FLG interface.[40,41] The 'thick region' with very weak photoluminescence signal corresponds spatially with the topographically thick region of the WSe$_2$ flake clearly visible in Figure 2. The peak of this emission is outside the range of the measurement (less than 1.5 eV), which is consistent with the indirect bandgap emission from bulk WSe$_2$.[17,42] In this photoluminescence map we also observe a weak broadband background emission from the Au substrate surrounding the sample, which we attribute to organic residue resulting from the transfer process since the PDMS makes direct contact with the substrate around the flake. Henceforth we direct our attention towards the 1L WSe$_2$/Au region where the emission is not quenched by substrate interactions.

Figure 4(c) shows the region of interest from the confocal photoluminescence map that corresponds with the AFM map in Figure 3(a). The spatial point spread function of this measurement, and the mapping step size are too large to resolve the wrinkles and nanobubble features, however we do observe some variation in the photoluminescence intensity across this area, indicative of inhomogeneity in the substrate surface and dielectric environment.[43,44] This variation is exemplified by extracting spectra measured at three different points labelled (i) to (iii), which are compared in Figure 4(d). These spectra exhibit a factor-of-1.7 variation in photoluminescence intensity but no significant variation in the shape or position of the emission (see normalised spectra inset).

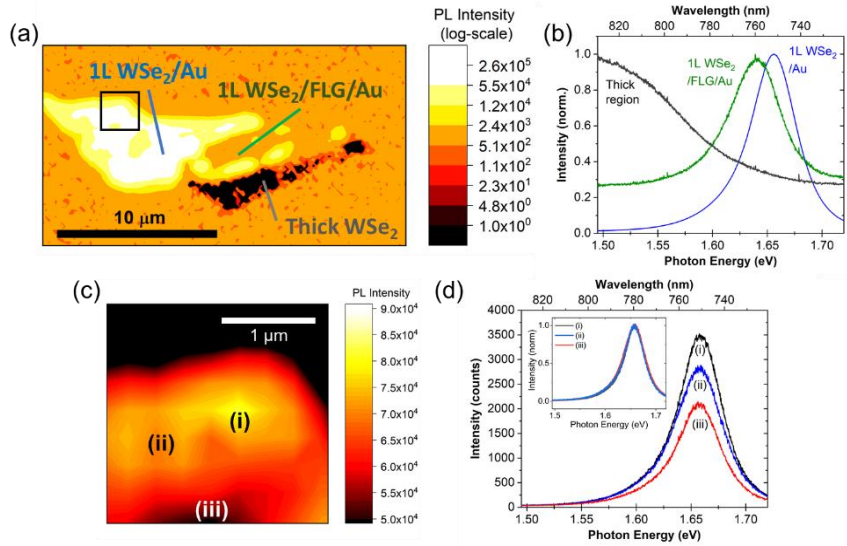

*Figure 4. (a) Confocal photoluminescence (PL) map of the sample with total integrated photoluminescence emission plotted as a colour map on a logarithmic scale. The 1L WSe$_2$/Au and 1L WSe$_2$/FLG/Au regions are indicated as well as a region of thicker WSe$_2$. (b) PL spectra (normalised to their peak) corresponding to spatial averages for regions indicated in (a). (c) Confocal PL map for the region of interest indicated by the black square in (a). (d) Example PL spectra extracted from positions (i)-(iii) indicated in (c) including inset with normalised spectra.*

In order to probe the optoelectronic properties of the nanoscale features in the 1L WSe$_2$ flake, we make use of tip-enhanced near-field spectroscopy to examine the region of interest. The measurement technique distinguishes the near-field from the far-field contribution as described in the Experimental Methods section. The TEPL results are presented in Figure 5. The total integrated near-field photoluminescence signal map in Figure 5(a) clearly resolves the pattern of topographic wrinkles with the strongest intensities corresponding to the crests of the wrinkles.

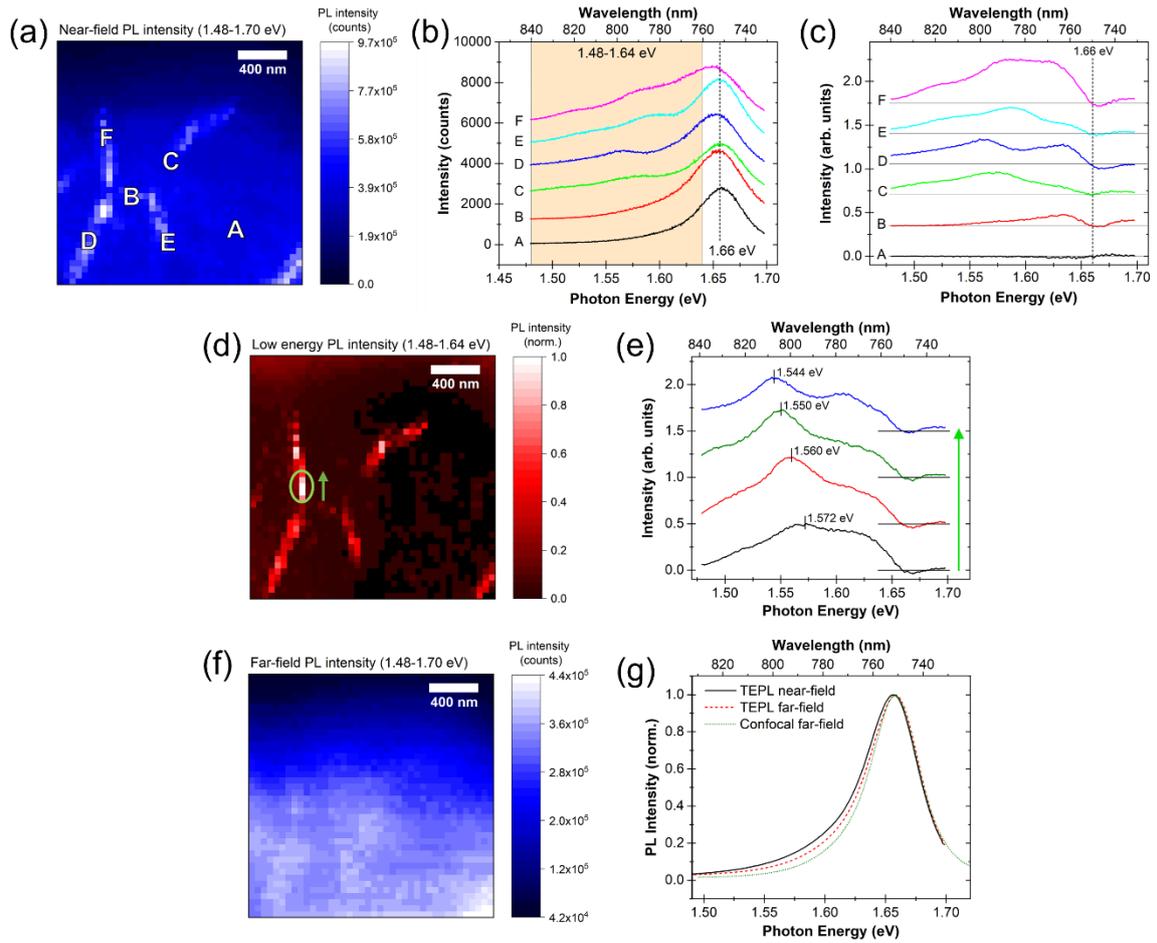

*Figure 5. Tip-Enhanced Photoluminescence (TEPL) mapping for the 1L WSe$_2$/Au region corresponding to the area indicated in Figure 3(a) with 40 pixels per line and 50 nm pixel size. (a) Map of integrated TEPL signal (1.48 – 1.70 eV) showing the near-field contribution, showing enhanced emission from wrinkles with selected positions indicated with letters A-F, matching Figure 3(a). (b) TEPL spectra measured at positions A-F with offsets in intensity for clarity. (c) TEPL spectra of localised emission (following normalisation to the intensity at 1.656 eV and subtraction of the averaged background emission) for positions A-F with offset in intensity for clarity. (d) Map of (normalised) integrated localised emission TEPL signal between 1.48 eV and 1.64 eV (after subtraction of the 1.656 eV background emission). (e) Comparison of localised emission spectra for four adjacent points with strong emission indicated by the green oval and arrow in (d) with spectra offset vertically for clarity. (f) Map of integrated signal (1.48 – 1.70 eV) showing the far-field signal acquired in the TEPL measurement, (g) Comparison of normalised spectra taken as spatial averages for the whole TEPL map (both near-field and far-field contributions) and for the same region of the standard confocal far-field PL map.*

The integrated photoluminescence intensity measured for the flat $WSe_2$ background (*e.g.* position A) is around $4 \times 10^5$ counts. For the tops of the wrinkles we measure intensities up to $1 \times 10^6$ counts, though with significant variation at different positions. To understand the nature of this overall enhancement in photoluminescence, we compare the near-field photoluminescence spectra for the selected points A-F in Figure 5 (b). At control position A (flat area), the near field emission shows a single peak centred at 1.656 eV, whereas the spectra for the wrinkles (positions B-F) exhibit strong additional contributions at lower energies down to 1.48 eV, whereas the higher energy peak shows minimal variation in spectral position over the range 1.653 eV to 1.658 eV. The localised emission at energies below the main bright exciton is shown more clearly in Figure 5 (c) where the background emission has been subtracted. This was achieved by normalising all the individual spectra in the map to the intensity at 1.656 eV and subtracting a spatially averaged background spectrum taken from the area surrounding point A (also normalised at 1.656 eV). The effectiveness of the background subtraction is demonstrated by the resulting flatness of the spectrum for point A. In contrast, the spectra taken from the wrinkles all show low-energy contributions, with point F showing the strongest emission across a broad range of energies.

The localised emission spectra associated with the 1L $WSe_2$ wrinkles show evidence of many contributing emissive states with significant point-to-point variation. The peaks of these emissive states span a range from around 1.52 eV to 1.63 eV. Attempts at fitting the full set of spectra from the map require a large number of overlapping emissive species, indicating that there is a semi-continuous distribution of emissive localised states within the wrinkled 1L $WSe_2$. Similar emission signatures have been observed in several other studies where they are identified as strain-localised excitons or defect states.[7,14] Since the TEPL measurement probes a small region of the sample (~10 nm diameter) the observation of a distribution of localized states gives some indication of the spatial extent and density of these states. The spatial resolution achieved in TEOS is the subject of much discussion in literature, but a resolution of around 10 nm has become an accepted value to quote in the general case.[32] Our use of gap-mode enhancement might be expected to achieve higher resolution than this, but would only be supported on the flat material since the wrinkled WSe$_2$ is delaminated from the substrate resulting in a larger tip-substrate separation. In practice the spatial resolution is highly variable and should be evaluated from features within each measurement dataset. The FWHM of the TEOS signal profiles across wrinkles is around 40 nm (see Supporting Information), but rigorously the spatial resolution can only be claimed to be < 150 nm (<3 pixels) which is less informative.[45] We return to the identification of these emissive states detected by the TEOS measurements in the Discussion Section.

The spatial distribution of the low-energy localised emission (integrating the range 1.48 eV to 1.64 eV after subtraction of the averaged background emission with the 1.656 eV peak) matches that of the total PL intensity (see Figure 5(d)), showing that this emission is characteristic of the wrinkled material. We note that there is no clear dependence on the orientation of the wrinkle, and that the strongest emission arises from isolated positions (represented by the white-coloured pixels occurring in clusters of 1-4 adjacent pixels) located along the crests of the wrinkles. Figure 5(e) compares the localised emission spectra for 4 adjacent pixels showing strong emission, we find that the spectra are similar in form with at least 2 spectral contributions but the strongest emission peak position varies between 1.544 eV and 1.572 eV (a range of 28 meV). Since each pixel has spatial size of 50 nm, this indicates that the energetic landscape experienced by the emissive species varies continuously on the length-scale of 100 nm to 200 nm along the wrinkle crest. We therefore deduce that the low-energy localised emission is characteristic of the wrinkled structure of the 1L WSe$_2$ and arises from a manifold of excitonic states or defect states, with binding energies that are tuned by the local structure of the wrinkle.

It is also important to compare the near-field emission with that observed in the far-field. In this study we have access to two distinct far-field measurements: the standard confocal spectroscopy shown in Figure 4 (obtained with the conventional top-illumination LabRam instrument), and the far-field signal from the TEPL measurement (obtained with the TRIOS side-illumination system in the Dual-Spec mode). The latter far-field photoluminescence map is presented in Figure 5(f), and the low spatial resolution means that the wrinkle features are not resolved. Due to the difference in spatial resolutions, comparison of the far-field and near-field spectra is achieved by spatially averaging spectra for the different measurements over the same area of interest. These spectra are presented (normalised to aid comparison) in Figure 5(g). All three spectra exhibit the same main bright exciton emission peak centred at 1.656 eV, however, the TEPL spectra show a broad tail on the low-energy side of the peak that is much weaker in the confocal far-field photoluminescence spectrum. The low-energy emission is strongest in the TEPL near-field spectrum, with the TEPL far-field spectrum representing an intermediate case. This discrepancy indicates that the localised emission is not accessible in top-illumination (in-plane polarisation) far-field spectroscopy. Whilst it is most strongly observed in the near-field measurement, the localised emission is accessible in the far-field using the side-illumination geometry (out-of-plane polarisation).[46]

The enhanced near-field photoluminescence in Figure 5 is clearly associated with the wrinkles in the 1L WSe$_2$ flake, where local strain or structural defects in the material are understood to play a role.[7,11,14] Such effects would modify the local phonon modes of the 1L WSe$_2$ and hence can be probed with Raman spectroscopy.[47,48] Using the same instrumentation as for TEPL but selecting a different spectral range, we are able to perform TERS to resolve these phonon signatures with nanoscale spatial resolution. Raman spectra from 1L WSe$_2$ are characterised by a complicated band of scattering around 250 cm$^{-1}$, where a number of modes have overlapping energies resulting in some debate over precise identifications.[49-51] The E' and A'$_1$ modes are degenerate at the Γ point of the Brillouin zone resulting in a combined peak typically reported at around 250 cm$^{-1}$.[49] However, since the 1.96 eV laser excitation stimulates a resonant scattering process, overtone and combination bands are allowed with strong contributions arising from phonons at the M and K points of the Brillouin zone, some of which also overlap with the modes at around 250 cm$^{-1}$.[51] For the TERS measurement we simply consider the integrated intensity over the main Raman scattering band with its peak at around 260 cm$^{-1}$. This is plotted as a map in Figure 6(a) showing a weak intensity from the flat background of the 1L WSe$_2$ flake, with linear features of high Raman scattering intensity. These features correspond with the topographic wrinkles observed above and hence reveal a local enhancement of the Raman scattering compared with that for the flat regions.

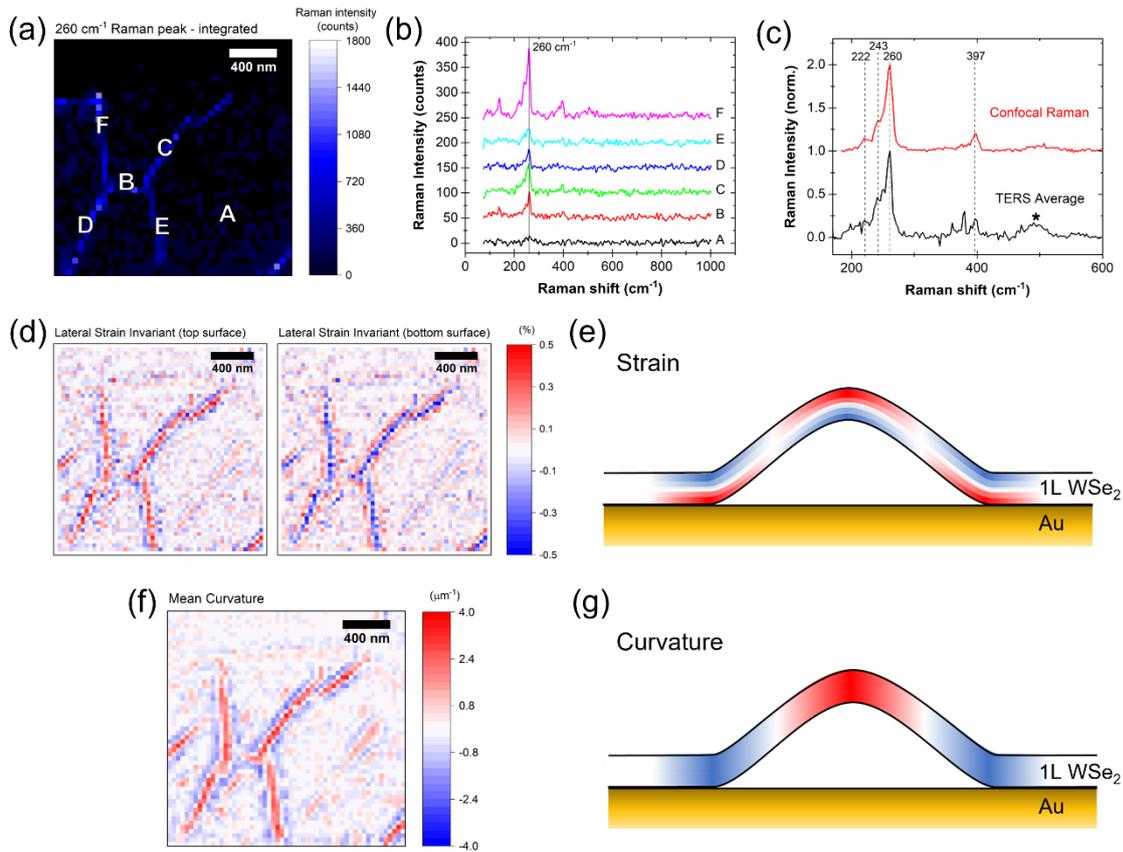

*Figure 6. (a) Map of integrated TERS signal for 260 cm$^{-1}$ peak for the 1L WSe$_2$/Au region with specific positions indicated with letters A-F corresponding to the area indicated in Figure 3(a). (b) TERS spectra measured at positions A-F with offsets in intensity for clarity. (c) Comparison of normalised Raman spectra taken as spatial averages over the whole TEPL map and for the same region using confocal Raman microscopy mapping with some distinctive Raman modes indicated and an \* indicating the additional 493 cm$^{-1}$ peak in the TERS spectrum. (d) Heat maps of lateral strain invariant for the top (left) and bottom (right) surfaces of the 1L WSe$_2$ extracted from topography map (Figure 3(a)). (e) Sketched cross-section of 1L WSe$_2$ wrinkle illustrating strain model with neutral strain through centre of the sheet and red/blue shading indicating regions of positive/negative lateral strain. (f) Heat map of mean curvature calculated for each point on the surface based on topography map. (g) Sketched cross-section of 1L WSe$_2$ wrinkle illustrating curvature model with red/blue shading indicating regions of high positive/negative curvature.*

To understand the local properties of the 1L WSe$_2$ wrinkles and the observed enhancement in Raman scattering intensity, we compare example TERS spectra for the points A-F in Figure 6(b). Aside from the variation in intensity of the Raman scattering, there are no discernible differences in the peak positions, shapes, or relative intensities. All spectra show the main peak at around 260 cm$^{-1}$ with a clear shoulder at the low-wavenumber side. The strongest spectrum, measured at point F, also shows modes centred at around 140 cm$^{-1}$ and 397 cm$^{-1}$, which are reported elsewhere as the first and third order LA(M) modes.[52] These weaker modes may also be present in the other spectra but are too weak to unambiguously observe. Achieving high signal-to-noise ratios in the TERS experiment is a particular challenge arising from the extreme confinement of the electric field and the need to use low excitation laser power to avoid damage to the tip or sample. However, the overall TERS spectrum can be compared with the far-field confocal spectrum by averaging the contributions over the full region of interest – this comparison is presented in Figure 6(c). The confocal Raman and TERS spectra look very similar: the confocal spectrum has several clear features that are identified using assignments from literature: 222 cm$^{-1}$ [TA(M)+ZA(M)], 242 cm$^{-1}$ [LA(M)+ZA(M)], 260 cm$^{-1}$ [2LA(M)], and 397 cm$^{-1}$ [3LA(M) or A1'(M)+LA(M)].[51] All of these modes are consistent with the features observed in the TERS spectrum. We do not see additional Raman modes associated with symmetry-breaking reported elsewhere,[53,54] though we note that other reports consider situations where the material is strongly coupled to the substrate, and we have shown that not to be the case here. The only significant difference is the appearance of a broad Raman band at around 493 cm$^{-1}$ in the TERS spectrum, which is barely visible or absent in the confocal spectrum. The appearance of such a Raman peak is typical of a resonant interaction, where the optical excitation is resonant with an electronic transition that is coupled to a particular phonon mode.[55,56] In this case, the fact that the peak is enhanced in the TERS spectrum compared with the confocal spectrum indicates that the *z*-polarised near-field couples to an electronic transition that is not accessible in the far-field measurement. This suggests that there is a dark state with an out-of-plane transition dipole at an energy close to 1.96 eV, corresponding to the He-Ne laser excitation used here. A Raman peak in this position was also reported by McDonnell *et al.* using excitation close to resonance with the A-exciton 2s Rydberg state at around 1.9 eV.[50] Their analysis of this feature similarly concluded that one or more dark states are required at this energy to explain its dispersion behaviour.

It is widely reported that the Raman peaks of TMD materials are sensitive to strain such that the peak positions (particularly the E' peak) can be used to determine the local strain. However, for $WSe_2$ the relevant phonon modes are nearly degenerate, making it difficult to resolve the small shifts of only a few cm$^{-1}$ for uniaxial strains of up to 3 %, which is close to the resolution limit of the instrumentation.[47,49,57] For the results in Figure 6(a) and (b) we observe no evidence of such strain associated with the wrinkles but the weak signal strength prevents satisfactory peak fitting that would be required to resolve the splitting of the modes around 260 cm$^{-1}$. However, the topographic information is also able to yield an estimate of the local strain in the wrinkles.

In general, the deflection of a plate under strain is described by the Föppl von Kármán equations, which have been applied to curved and wrinkled sheets.[38,58-60] However, this approach requires the use of boundary conditions or symmetry considerations that do not apply directly to the case of a free-standing wrinkle. Specifically, we are unable to evaluate the unstrained extent of the material without additional knowledge. An alternative approach to evaluating strain in material with a buckling delamination balances the bending energy with adhesion (interfacial toughness), but assumes that the in-plane strain is zero.[23,37,61] Rather than introduce additional unknown parameters, we have adopted a simplistic (but nevertheless instructive) approach to estimate the strains based only on the topography data, as described in the Experimental Methods. To do this we also assume that the $WSe_2$ has a thickness of 0.65 nm with a zero-strain plane running through the centre of that thickness.[62] The resulting lateral strains experienced by the top and bottom surfaces are plotted in Figure 6(d) alongside a sketch illustrating the interpretation of this approach in Figure 6(e). For the top surface we find that the crests of the wrinkles have a maximum tensile strain of 0.5 %, whereas the sides of the wrinkles show a maximum of compressive strain with a slightly smaller magnitude where the wrinkle approaches the substrate. Note that the bottom surface experiences the inverse strain in this case. We note that this calculation uses a small-deflection approximation, which is not well-matched by the large (with respect to the monolayer thickness) wrinkles in our samples, however, this does not affect the qualitative strain distribution, and the strain values obtained of <1 % are consistent with other reported values in the literature for similar studies.[11,19,58] We note that these other studies did not generate or calculate strain in a comparable way, and indeed we propose that curvature is a more appropriate consideration in our case. Whilst Figure 6(d) plots lateral strain values for the whole region of interest, it is important to recognise that the 'free plate' assumption is only expected to apply to the wrinkled areas since contact forces between the 1L $WSe_2$ and the substrate can cause lateral strains that are not represented by the surface topography.[63-65]

Much larger values for strain in curved 2D materials have been reported elsewhere, so it is important to understand the differences. In large part, these arise from our understanding of the nature and origin of the wrinkles: we consider the wrinkles as buckling delaminations formed by compressive strain arising from the transfer process, where the in-plane strain is expected to be negligible within the delaminated regions.[59]

As illustrated by Figure 6(e) the strain field in a wrinkled monolayer cannot be understood in terms of simple in-plane tensile or compressive strains. A more intuitive model for such wrinkles is to consider the curvature. Since the wrinkles have different orientations, the mean curvature at each position is the most helpful parameter to consider and can be understood as the inverse of the radius of curvature. These values are plotted as a spatial map in Figure 6(f) alongside a sketch illustrating the interpretation of this approach in Figure 6(g). The result may appear qualitatively similar to the strain map in Figure 6(d) but is conceptually distinct. The mean curvature is greatest at the wrinkle crests, and the sides of the wrinkles have a smaller magnitude of curvature with an inverted sign. We observe that the maximum curvature at the wrinkle crests corresponds well with the areas of enhanced PL and Raman emission in the tip-enhanced measurements.

**Discussion**

Considering all the results together allows us to address two important questions more comprehensively:

(i) What is the nature of the low-energy localised emissive states associated with wrinkles that are detected in the TEPL measurement?

(ii) What is the mechanism for local enhancement of the PL emission and Raman scattering signals in the near-field spectroscopy measurements at the wrinkle crests?

We address these questions in turn.

*(i)     Nature of the low-energy localised emissive states associated with wrinkles*

The photoluminescence spectra presented in Figure 5 show strong emission from the 1L $WSe_2$ free bright exciton at 1.66 eV as well as contributions from a semi-discretised broad band of lower energy emissive states including spectral peaks and shoulders across the range from 1.52 eV to 1.63 eV. The free bright excitonic emission is observed in both the confocal (far-field) and tip-enhanced (near-field) measurements. In the TEPL measurement, which offers high spatial resolution, we also find that this emission is similarly observed from the flat and wrinkled regions. Conversely, the low-energy emission is only observed in the near-field measurement and is spatially localised to the wrinkles. These observations lead us to conclude that the combination of both the near-field measurement regime and wrinkled nanostructure is required to stimulate and detect the low-energy emission.

The identification of the free bright exciton at 1.66 eV (the A-exciton) is well-attested in published studies. Several other excitonic species have also been reported at lower energies, including trions, biexcitons, and dark excitons.[14,66] These species have energies up to 50 meV below the free neutral exciton, whereas the emission we observe exhibits features in the range 30 meV to 150 meV below the main excitonic peak and so could only be partially accounted for by such excitonic species. The highest energy feature (1.631 eV) we detect could be identified as the dark exciton, which was observed at 1.62 eV by Park *et al.* in a similar TEPL measurement for flat 1L WSe$_2$ on a gold substrate.[34] The latter study is pertinent because the authors demonstrate that the spin-forbidden dark exciton can be excited using the gap-mode tip-enhancement, where the *z*-polarisation of the surface plasmon polariton couples to the out-of-plane dipole of the dark exciton. In their study the emission is detected from flat 1L WSe$_2$ and the Purcell effect is also invoked to explain the enhanced emission. The intrinsic spin-flip required for radiative decay of this dark exciton is facilitated by the broken mirror symmetry in the *z*-direction corresponding to a monolayer interacting with the substrate. Theoretical calculations of curvature in monolayer TMDCs show the emergence of an effective in-plane magnetic field and a Rashba-like spin-orbit coupling that would similarly enable intrinsic spin flipping in the wrinkled morphology considered here.[67] However, as noted above this dark exciton is not able to account for the wide energy range of localised emissive states that we observe.

The energy range of emission that we observe is more consistent with the defect band from 1.5 eV to 1.7 eV reported by Parto *et al.* for electron-beam induced defects.[7] In our case, the high spatial resolution and sensitivity of the TEPL technique allows us to observe structure within the defect band emission rather than just a single broad peak. Tensile strain in the 1L WSe$_2$ is widely considered to play a role in explaining localised exciton emission with energies down to 1.565 eV [13] and 1.515 eV [7], corresponding to confinement potentials of 90-140 meV with respect to the free bright exciton. Some studies explain this in terms of strain-induced narrowing of the bandgap, but this effect alone would require a strain of several percent.[10,11,21,22] Such strain values can be achieved by various methods but, as discussed above (see Figure 6(c)), we would not expect this reasoning to apply for the free-standing wrinkles. The study by Koo *et al.* examined the shift in the free bright exciton emission for similar wrinkles in 1L WSe$_2$ and reported a modest shift of around 10 meV with a tensile strain of 0.2 %.[11]

The developing consensus in the field is that low energy localised emission (like that reported in this study) originates from defect states, which are ubiquitous in 1L WSe$_2$, and that strain plays a role in enhancing or tuning the emission from these states.[7,24,26] The results we have presented cannot be explained within this paradigm, since the wrinkles do not exhibit significant in-plane strain, and are more appropriately described in terms of curvature. In our case, we also observe that the localised photoluminescence is strongly coupled to the *z*-polarised surface plasmon polariton, this indicates that the transition dipoles associated with the optical absorption and/or emission are oriented out-of-plane. A recent report similarly shows predominantly out-of-plane emission from localised excitons in strained 1L WSe$_2$ for energies below 1.653 eV.[13] Drawing together this discussion, we conclude that the low energy emissive states we observe originate from defect states with out-of-plane dipoles. These states are spin-forbidden in flat (and unstrained) 1L WSe$_2$ but activated by the application of curvature that breaks the local mirror inversion symmetry and enables intrinsic spin flipping.[67,68]

  *(ii)    Mechanism of enhanced near-field PL and Raman scattering from wrinkles*

The above identification of the emissive states as defect states with out-of-plane dipoles that are activated by curvature of the 1L WSe$_2$ at the wrinkle explains why their localised emission is not observed for the flat material around the wrinkles. However, there are other possible mechanisms affecting the differences in intensity of the PL and Raman scattering signals measured for the wrinkles that should be considered.

Exciton funnelling is widely discussed in literature as a potential mechanism for enhancement of photoluminescence in strained 1L WSe$_2$. The reasoning proposed is that the strain-induced reduction in the direct bandgap leads to drift of excitons towards the strained region. This simplistic model is not able to explain the enhanced emission observed in our TEPL results:

Firstly, if we simply consider the crest of the wrinkle as having the highest strain and hence the narrowest direct bandgap then we would expect to see a substantial red shift of the local emission from the free neutral exciton (at 1.656 eV). However, we observe very little spectral shift (± 3 meV) in the main bright exciton emission from the wrinkle crests indicating minimal strain-induced bandgap narrowing. Instead we find the emergence of additional lower-energy localised emissive states without suppression of the bright exciton. The highly localised plasmonic tip-enhancement (~10 nm) indicates that the bright exciton and lower-energy emission both originate from the same spatial region, though we cannot fully exclude a contribution from stray far-field emission, and have already noted the difficulty in establishing the spatial resolution achieved in this measurement. Also, the wrinkle topography means that we might expect coupling between the z-polarised near-field enhancement and in-plane transition dipoles of the $WSe_2$ associated with the tilted sidewalls of the wrinkle, which could contribute to the strength of emission from the free neutral exciton.[69] Even if it were argued that the un-shifted exciton emission comes from unstrained material rather than the wrinkle apex itself, the exciton funnelling mechanism is unable to explain the out-of-plane transition dipole of the localised emission.

Secondly, our proposed understanding of the strain field in a free-standing wrinkle is not consistent with exciton funnelling, which requires in-plane tensile strain to reduce the direct bandgap.[21] A free-standing wrinkle cannot support overall in-plane tensile strain at the apex. This is in contrast to the case for nanobubbles, which are supported by hydrostatic pressure, and strained wrinkles formed on structured substrates.[10,12,58] In our case, it is also possible that there is trapped material under the wrinkles, noting particularly that not all the wrinkles extend to the edge of the flake and the possibility of trapped material even under an open-ended wrinkle. Another possibility is that the contact force of the AFM probe itself on the wrinkle results in a local distortion and associated strain, but this would again result in a local shift in the direct bandgap emission for the wrinkles with respect to the flat 1L $WSe_2$ supported on the substrate.[11] Alternatively wrinkles can form to accommodate strong tensile strain applied parallel to their long-axis, but this is not consistent with the topography reported in this paper since we would expect such wrinkles to show a strongly preferential alignment and might also expect to observe evidence of material extension parallel to the wrinkles.[70] The absence of detectable strain in the Raman spectroscopy measurements is most consistent with our proposed model showing minimal in-plane strain of the wrinkled material. Within the model of the wrinkles as buckling delaminations, it is possible for there to be a residual in-plane strain, but this would be compressive in nature and hence expected to result in widening of the bandgap.[21,71]

Another possible explanation for the local enhancement of PL for the wrinkles is that the physical separation of the 1L WSe$_2$ from the substrate might result in reduced quenching through the semiconductor-metal interaction.[72] We have already seen that the 1L WSe$_2$ is effectively decoupled from the Au substrate, which resulted in the strong emission from 1L WSe$_2$/Au regions in contrast to the 1L WSe$_2$/FLG/Au regions of the sample (see Figure 4(a)). Furthermore, both the gap-mode enhancement and Purcell effects would act in the opposite direction, tending to increase the tip-enhanced signal for the smaller tip-substrate distances associated with the flat regions of the sample.[34] We also note recently published work considering the differences between momentum-dark and bright exciton funnelling associated with local strain,[73] and a study on bilayer WSe$_2$/hBN nanobubbles showing that the dielectric environment (distinctly from local strain) can affect exciton funnelling.[74] Whilst these studies do not directly relate to samples studied here, their insights are relevant to the broader discussion of the mechanism of enhanced localised emisison in such materials.

Our proposed explanation for the enhanced TEPL signal at the wrinkles follows naturally from the identification of the emissive species as curvature-activated defect states with out-of-plane transition dipoles. These states are spin-forbidden for the flat material due to the preserved mirror symmetry and so only the free bright exciton with its in-plane transition dipole is observed, whereas for the curved 1L WSe$_2$ at the wrinkle the broken symmetry enables the primarily *z*-polarised tip-enhanced near-field to couple efficiently to the localised defect states with out-of-plane dipoles, as illustrated in Figure 7(b). Due to the difference in energy between the optical excitation (1.96 eV) and the energies of the emissive states (1.52 eV to 1.63 eV) we must also account for how these out-of-plane dipoles are created. The solution to this was revealed by the TERS results, which also show local enhancement for the wrinkled 1L WSe$_2$ compared with the flat regions. The TERS spectrum shows a scattering peak at 493 cm$^{-1}$ that is absent in the confocal Raman spectrum, and which evidences a resonant enhancement with a dark excitonic state close to 1.96 eV that is only accessible through the *z*-polarised near-field. We propose that this dark state is responsible for both resonant enhancement of the local Raman scattering and photoluminescence emission for the wrinkled 1L WSe$_2$. In the latter case, the dark exciton is directly excited by a 1.96 eV photon and non-radiatively relaxes to defect states, which also have out-of-plane transition dipoles (rather than to the in-plane polarised free bright exciton). It is the emission from these defect states that is detected in the TEPL measurement. This proposed mechanism is represented schematically in Figure 7(b).

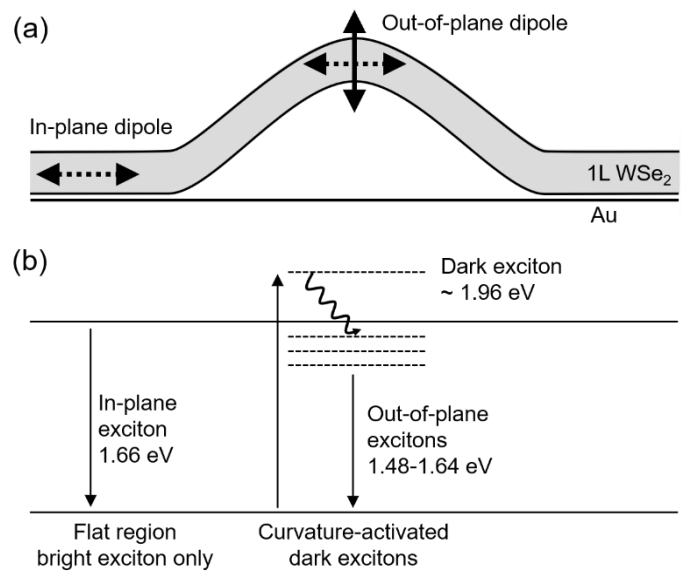

*Figure 7. Schematic summary of the proposed model for localised photoluminescence associated with 1L WSe$_2$ wrinkles. (a) Diagram showing cross-section of 1L WSe$_2$ wrinkle on Au substrate where the out-of-plane dipole (solid arrow) emission is active in the curved region the in-plane dipole (dashed arrow) emission is active in both the flat and curved regions. (b) Energy level sketch comparing the flat region, which only exhibits the 1.66 eV exciton with an in-plane dipole, with the*

*wrinkle, which exhibits curvature-activated dark excitons including a dark state near 1.96 eV and a manifold of lower energy defect states with characteristic out-of-plane emission dipoles.*

Whilst our proposed model has been developed specifically for the case of free-standing wrinkles, where the in-plane strains are minimal, it is valuable to consider how this might relate to two other instructive situations described in literature, namely pure in-plane stretching strain, and combined curvature with in-plane strain.

The case of in-plane uniaxial or biaxial strain is relatively easy to model, and Niehues *et al.* provide a relatively clean experimental verification.[75] It is clear that applied in-plane tensile strain causes a reduction in bandgap manifest by the shifting of the free bright exciton emission and absorption together towards lower energy. A strain sensitivity of around 50 meV/% is demonstrated for uniaxial strain in 1L $WSe_2$. Other studies indicate that a stronger effect can be achieved through biaxial strain, and that compressive strain causes a widening of the direct bandgap.[21,71] Importantly, this effect (for strains up to a few %, at least) modifies the energy of the A-exciton but does not change its nature. The transition dipole remains in-plane, and there is no evidence of activation of dark (spin-forbidden) states. Exciton funnelling can be easily understood in this context.

The case of nanobubbles (and similar straining strategies) combines in-plane tensile strains with curvature of the delaminated material and provides access to strongly localised emission. Compared with pure uniaxial strain these experiments show stronger shifts in the emission energy of 100 meV to 150 meV below the bandgap, with strains of around 1 %.[7,12,14,46] Several studies show that this emission has a distinct nature from the A-exciton requiring the proposed formation of new localised electronic states, emission from dark-excitonic states, defect sites, or the local dielectric environment.[12,13,34,76,77]

We have argued that free-standing wrinkles (buckling delaminations) provide an instructive intermediate case with strong curvature but minimal in-plane strain. We have shown phenomenologically similar localised emission to that from nanobubbles (discussed further in the Supporting Information) but, in the absence of strong in-plane strains, we have proposed curvature-induced symmetry breaking an alternative explanation. Furthermore, we suggest that this mechanism might be considered to help explain the strongly localised emission and dark-exciton activation observed in some of studies cited above where in-plane strain and curvature are hard to separate.

**Conclusions**

We have studied room-temperature localised emission from free-standing wrinkles in 1L $WSe_2$ using TEOS to spatially resolve the nanoscale structure and optoelectronic properties. We observed low energy localised emission with out-of-plane transition dipoles at the crests of the wrinkles, which we assigned to defect states. The experimental results and strain calculations presented cannot be accommodated within the dominant paradigm in published literature. Rather, we propose that curvature rather than strain of the 1L $WSe_2$ is the most appropriate explanation for these results. We have presented evidence for a manifold of dark excitonic states, through which excitation at 1.96 eV is coupled to low energy localised defect states.

The results of this study suggest curvature engineering of 1L $WSe_2$ as an effective route towards localised emission at high temperature, and an alternative strategy to the pursuit of increasing in-plane stretching strain to confine excitons. The spectral features we report at room temperature show energetic confinement comparable with reports of single photon emission in 1L $WSe_2$ so photon antibunching experiments are an interesting area for future study.[6,7]

It is significant that the emission we report here originates from optical coupling to spin-forbidden dark states, which are typically long-lived and therefore of potential interest for quantum information technologies.[34] The out-of-plane transition dipoles for this emission are also relevant for many proposed applications of TMDs for plasmonic circuits and on-chip photonic devices, which require the coupling of light into 'horizontal' waveguides.[13,25] This is in contrast to the typically-observed free bright excitons in TMDs, which have in-plane dipoles that efficiently couple light in/out of thin film devices structures in a 'vertical' direction. Furthermore, the localised emission we observe exhibits semi-continuous spatial variation in the spectral positions of the peaks so we speculate that it may be possible to tune the emission through the local curvature of the 1L WSe$_2$.

**Experimental Methods**

*Sample Preparation*

The template-stripped gold substrate was purchased commercially (Platypus Technologies). It consists of an ultra-flat 100 nm thick Au film attached with epoxy to a 1 cm × 1 cm aluminosilicate glass chip. The root-mean-square roughness is quoted as 3.6 Å. Few-layer graphene (FLG) was mechanically exfoliated from graphite via the scotch-tape method and transferred to a clean Au substrate. The WSe$_2$ monolayer was mechanically exfoliated from a bulk crystal (HQ graphene) onto a polydimethylsiloxane (PDMS) film. Using the viscoelastic stamping method, the monolayer WSe$_2$ was aligned and transferred onto the previously deposited FLG flake via a micrometre-precision transfer station. The sample was prepared in an ISO7 (class 10,000) cleanroom to minimise contamination.[78] An Olympus MX63 was used for optical microscopy.

*Atomic Force Microscopy (AFM)*

Atomic force microscopy (AFM) based topography measurements were carried out in non-contact mode with an AIST-NT (now HORIBA Scientific) Combiscope 1000. Au-coated probes with resonant frequency of ~70 kHz and nominal spring constant of 2 N/m were used (MikroMasch, OPUS 240AC-GG). Median line correction and plane correction were applied to the measured data as well as tear correction using adjacent averaging. Processing and plotting were performed using Gwyddion software.[79]

*Confocal Photoluminescence (PL) and Raman Spectroscopy*

Confocal PL and Raman spectroscopy was carried out using a confocal microspectrometer (HORIBA LabRam HR Evolution) using a 100×, 0.9 NA objective. A 1.96 eV (633 nm) He-Ne pump laser was used for excitation and the collected signal was dispersed using a 300 grooves/mm grating achieving a resolution of around 2.5 cm$^{-1}$. This configuration was chosen to enable both Raman and PL spectroscopy with the same measurement setup. For confocal PL, the laser power was set at 500 µW, with an acquisition time of 0.1 s per pixel. For Raman spectroscopy, the laser power was set to 1 mW with an acquisition time of 6 s per pixel. Confocal Raman spectra were background-corrected using a high-order polynomial fitting line to subtract the photoluminescence contribution. References to standard confocal spectroscopy refer to the use of this system in top-illumination mode, where the optical path is normal to the sample plane.

*Tip-Enhanced Photoluminescence (TEPL) and Tip-Enhanced Raman Spectroscopy (TERS)*

TEPL and TERS were performed using the AIST-NT (now HORIBA Scientific) Combiscope 1000 instrument in a TRIOS system, which provides side-illumination optical access at an angle of 65 ° from the *z*-axis. Plasmonically active probes for both techniques were fabricated by coating commercial silicon probes with a layer of silver. Force-modulation type silicon probes (NANOSENSORS ATEC-FM) were used with a nominal resonant frequency and spring constant of 85 kHz and 2.8 N/m respectively. A silicon oxide layer was grown by baking for 45 minutes inside a tube furnace at 1000 °C under continuous water vapour flow. Organic contaminants were removed by an UV-ozone cleaner (UVOCS T10X10/OES/E) operating for 45 minutes prior to metal deposition.

The oxidised probes were mounted inside the thermal evaporator (MBRAUN LABmaster SP/DP) using a custom-made probe holder to maintain the optimised angle of the AFM tip axis to the evaporation source. A 75 nm thick layer of Ag was deposited onto the probes from a high purity Ag wire (Agar Scientific) at an evaporation rate of 0.1 nm/s under ~2 × 10$^{-7}$ mbar vacuum conditions. On top of the silver layer, a 2 nm thick aluminium layer was deposited from high purity Al wire (Agar Scientific) at a rate of 0.05 nm/s. The aluminium layer oxidises into AlO$_x$, which acts as a protective barrier to decelerate the degradation of the underlying plasmonic silver during tip-enhanced operation.[80,81] Whilst the resulting tip radius is relatively large for AFM, the strong plasmonic enhancement results in a substantially smaller spatial resolution for the tip-enhanced spectroscopy.

The plasmonic probes were excited with the He-Ne laser coupled through a 0.7 NA, 100× infinity corrected long working distance objective (Mitutoyo, Plan Apo). These measurements were performed in a 'side illumination' geometry where the objective is positioned at ~60 ° from the surface normal of the sample.

TERS and TEPL signals were collected in backscattering configuration and coupled into the HORIBA LabRam spectrometer (see details above). TEPL spectra were recorded with around 100 µW laser power with 100 ms acquisition time and using Dual-Spec operation where, at each point of the raster scan, spectra are acquired both with the probe held using tapping mode feedback and with contact feedback. Due to the strong distance dependent decay of the near-field, the spectrum acquired in tapping mode corresponds to the far-field measurement, which is then subtracted from the contact mode measurement channel to isolate the near-field. TERS spectra were acquired with the same laser power as for TEPL but with 500 ms acquisition time per pixel recording a single spectrum in contact mode at each point of the raster scan. These measurements (and all the others in this study) were conducted at room temperature in ambient conditions.

TERS and TEPL spectra were smoothed with 5-point moving average filter. The individual TERS spectra presented were background-corrected by manually fitting cubic splines to remove the photoluminescence contribution. To subtract the TEPL background signal a spatially averaged spectrum from a flat region of the sample was taken. All the spectra were normalised at the peak position of 1.656 eV, and the background signal was subtracted from each spectrum. Due to the complexity of the spectra preventing consistent fitting routines, spectral peak positions were simply read-off from the intensity maxima of the smoothed spectra. The Raman peak positions given in the manuscript were extracted from the confocal Raman spectrum (since it has stronger signal) by fitting with gaussian peaks.

*Strain Calculations*

In-plane strains on the top surface of the flake, caused by local bending, were calculated from the variation in the AFM height maps.[82] Map pixels were modelled as cuboidal elements (32 × 32 × 0.65 nm³) corresponding to the map pixel size and a flake thickness ($t$) of 0.65 nm.[62] Local out-of-plane rotation angles $\omega_{zx}$ and $\omega_{zy}$ were calculated numerically by taking the second derivative of the heights (*i.e.* the local curvature in *x* and *y*) and multiplying by the pixel step sizes $\Delta x$ and $\Delta y$ respectively.

$$\omega_{zx} = \frac{\partial^2 z}{\partial x^2} \cdot \Delta x \quad \text{and} \quad \omega_{zy} = \frac{\partial^2 z}{\partial y^2} \cdot \Delta y$$

The local rotation of the flake is accommodated by deformation of the cuboidal pixel element. The neutral axis in bending is assumed to be in the mid-plane of the flake, so that the top surface is in tension, and the bottom surface is in compression. This is a good physical approximation near the wrinkle peaks, where the bottom surface of the flake is not subjected to frictional forces from the substrate. The rotation angles are related to the surface tensile strains ($\epsilon_{xx}$ and $\epsilon_{yy}$) geometrically. This analysis assumes that the monolayer WSe₂ is isotropic in-plane.

$$\tan\frac{\omega_{zx}}{2} = \frac{\epsilon_{xx}}{2t} \quad \text{and} \quad \tan\frac{\omega_{zy}}{2} = \frac{\epsilon_{yy}}{2t}$$

The local tensile strain components were summed and presented as a map of scalar invariant lateral strain ($\epsilon_{xx} + \epsilon_{yy}$), we note that $\epsilon_{zz}$ (the out-of-plane strain) cannot be evaluated here and is neglected.

Curvature for each point was calculated for each point on the surface by evaluating the first and second fundamental forms and extracting the two principal curvatures. The mean curvature is the average of curvature over all directions in the *x-y* plane.[67,83] Calculations were performed using MATLAB.[84]

**Competing Interests**

The authors declare no competing interests.

**Supporting Information**

Supporting Information contains: comparison of TEPL spectra for nanobubbles and wrinkles, line profile comparison of topography and TERS signal.

**Author Contributions**

S.W., F.R., O.K., and F.A.C devised and planned the work. A.C. prepared the sample and contributed the optical microscopy. F.R. and S.W. contributed the optical spectroscopy and atomic force microscopy measurements and analysis of data. V.T. and S.W. contributed the strain analysis. S.W., F.R., T.V., V.T., Y.C., O.K., and F.A.C contributed to the interpretation of the results. All authors contributed to the technical discussion and preparation of the manuscript.

**Acknowledgements**

This work received funding from the UK Department for Science, Innovation and Technology (DSIT) through the National Measurement System. This project has received funding from the European Union's Horizon 2020 Research and Innovation programme under grant agreement GrapheneCore3 881603, and under the Marie Skłodowska-Curie grant agreement 721874 (SPM2.0).

**Table of Contents**

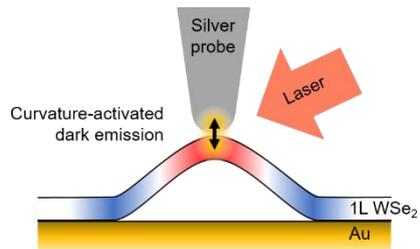

Localised emission from defect states in monolayer transition metal dichalcogenides is of great interest for optoelectronic and quantum device applications. We demonstrate highly localised, out-of-plane polarised photoluminescence from a tungsten diselenide wrinkle at room temperature using tip-enhanced plasmonic coupling. Detailed analysis with nano-resolution spectroscopy reveals a manifold of dark states that are activated by curvature.

# Supporting Information

# Curvature-enhanced localised emission from dark states in wrinkled monolayer WSe$_2$ at room temperature

Sebastian Wood*, Filipe Richheimer, Tom Vincent, Vivian Tong, Alessandro Catanzaro, Yameng Cao, Olga Kazakova, and Fernando A. Castro

*National Physical Laboratory, Hampton Road, Teddington, Middlesex, TW11 0LW, UK*

*Corresponding author: sebastian.wood@npl.co.uk

**Comparison of tip-enhanced photoluminescence spectra for nanobubbles and wrinkles**

The manuscript proposes that wrinkles in 1L WSe$_2$ are distinct from nanobubbles, due to their difference in structure and the associated strain field. Whilst the study has focused on the properties of the wrinkles themselves, the presence of nanobubbles visible in the topographic AFM presents an opportunity for a direct comparison. Figure S8(a) shows the AFM topography measurement area of interest highlighted in the main manuscript Figure 3(a), with the addition of crosses marking specific points to compare: point A is an example of a wrinkle, point B is a flat background point, whereas points 1-4 correspond with various nanobubbles identified in the topography channel. Figure S8(b) shows the TEPL near-field spectra extracted for the same points of interest (registration limited by the 50 nm pixel size in the latter case). The same spectra are shown normalised in Figure S8(c) to aid comparison.

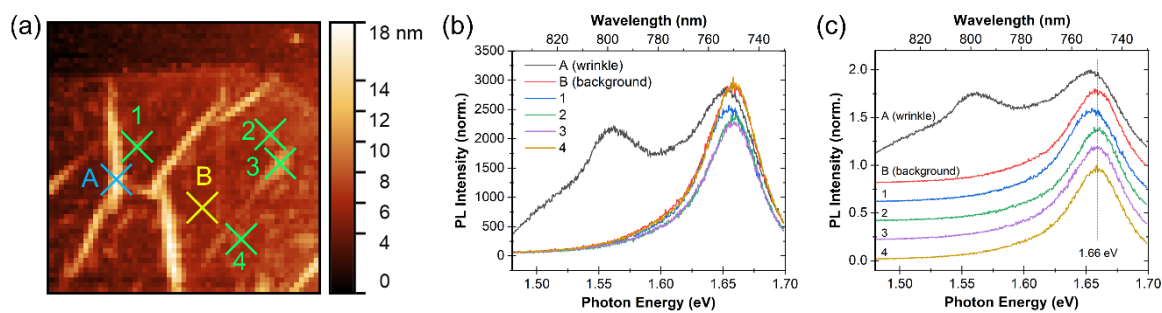

*Figure S8. (a) Topography AFM (left) AFM topography with crosses marking points of interest for comparison: A = nanowrinkle, B = flat background, 1-4 = nanobubbles. (centre) TEPL spectra corresponding to the points of interest. (right) Normalised TEPL spectra with vertical offsets for clarity.*

Comparing the spectra, we find that those corresponding with the nanobubbles (1-4) show very little difference from the flat background spectrum (B). In all cases the spectra are dominated by the 1.66 eV main bright exciton. Spectra 1,2, and 3 show a slightly reduced overall intensity, and spectrum 1 shows a red-shift of the emission peak (~ 5 meV), which may be indicative of strain. However, none of the nanobubble spectra show the low energy emission comparable with the wrinkle (spectrum A). This outcome is consistent with the proposed difference between nanobubbles and wrinkles outlined in this manuscript.

In comparison with other reports of localised emission from nanobubbles in literature, we note that the topographic features we present above appear similar to those reported elsewhere and so it is interesting that we do not observe similar photoluminescence signatures. Furthermore, the low-energy emission signatures we observe from the wrinkles are similar to those observed elsewhere for nanobubbles. Rodriguez *et al.*, provide a good comparison with emission around 1.55 eV from a WSe$_2$ nanobubble, which they associate with a strain of around 1 %. [1]. Clearly our measurement is sensitive to such emission so the fact that we do not observe it for the nanobubbles indicates a difference in the sample itself. The most likely explanation is the difference in substrate: the comparable nanobubble reports we cite use hBN as a substrate, [1] [2] [3] [4] whereas we have used Au. It is known that the substrate material can have a profound effect on the energy and intensity of nanobubble emission. [5]

A side-by-side comparison of our samples with those from other groups could enable further insight into the differences.

**Comparison of line profiles of topography and tip-enhanced Raman spectroscopy signal**

Estimation of the spatial resolution achieved in tip-enhanced optical spectroscopy is discussed in the manuscript. In principle this can be determined from the dilation of narrow features in the experimental data set. In this case the line-profiles measured (approximately) perpendicular to the wrinkles provides a measure of this. Figure S2 shows line-profiles across two wrinkles ('F' and 'C' in manuscript Figure 3(a)). The measured FWHM of the topography profile is 85-86 nm for these, whilst the TERS signal profile FWHM is 40-44 nm. The width of the topography profile is dilated by the physical sharpness of the probe tip and so provides an upper-bound for the wrinkle width. The TERS profile appears to be narrower, suggesting that the enhanced signal corresponds only with the apex of the wrinkle. This could be taken as an upper-bound for the spatial resolution achieved in the TEOS measurement, but in fact this is not robust since there are insufficient data points measured across the profile. A minimum of 4 points (3 pixels) is required so, with the 50 nm pixel spacing, we can only robustly claim resolution of < 150 nm. [6]

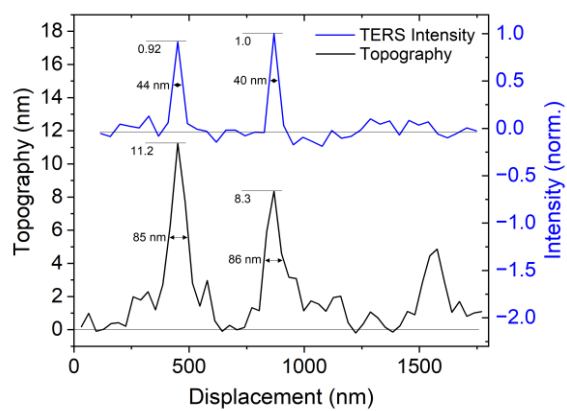

*Figure S2. Comparison of line profiles in topography and TERS intensity extracted from a horizontal line across the centre of manuscript Figure 6(a) and corresponding line in Figure 3(a).*